\documentclass[letter]{aa} 

\usepackage{graphicx}
\usepackage{txfonts}
\newcommand{\xmm}{{\em XMM-Newton}}

\newcommand{{\myr}}{\mbox{[$M_\odot\,{\rm yr}^{-1}$}]}
\newcommand{\lsim}{\raisebox{-.4ex}{$\stackrel{<}{\scriptstyle \sim}$}}
\newcommand{\msim}{\raisebox{-.4ex}{$\stackrel{>}{\scriptstyle \sim}$}}

\newcommand{\flux}{erg\,cm$^{-2}$\,s$^{-1}$}
\newcommand{\new}{ }
\newcommand{\nnew}{}
\newcommand{\cas}{IRAS\,00500+6713}

\begin{document} 


\title{X-rays observations of a super-Chandrasekhar object reveal an ONe and a CO white dwarf merger 
product embedded in a putative SN\,Iax remnant 
\thanks{Based  on the observations with the id numbers 0841640101, 0841640201 obtained with the ESA science  
mission \xmm.}}

\authorrunning{Oskinova et\,al.}
\titlerunning{X-rays observations of a super-Chandrasekhar object reveal an ONe and a CO white dwarf 
merger}

\author
{
Lidia M. Oskinova\inst{1,2}, Vasilii V. Gvaramadze\inst{3,4,5}, G\"otz Gr\"afener\inst{6}, 
Norbert Langer\inst{6,7}, Helge Todt\inst{1}  
}
\institute{\inst{1}{Institute of Physics and Astronomy, University of Potsdam, 
14476 Potsdam, Germany}\\
\email{lida@astro.physik.uni-potsdam.de}  \\    
\inst{2}{Department of Astronomy, Kazan Federal University, Kremlevskaya Str 18, Kazan, Russia}\\
\inst{3}{Sternberg Astronomical Institute, Lomonosov Moscow State University, Moscow 119234, Russia}\\
\inst{4}{Space Research Institute, Russian Academy of Sciences, Moscow 117997, Russia}\\
\inst{5}{Evgeni Kharadze Georgian National Astrophysical Observatory, Abastumani, 0301, Georgia}\\
\inst{6}{Argelander-Institut f\"ur Astronomie, Universit\"at Bonn, Germany}\\
\inst{7}{Max-Planck-Institut f\"ur Radioastronomie, Bonn, Germany}
 \date{Received ? / Accepted ?}
}

 
\abstract{
The merger of two white dwarfs (WD) is a natural outcome from the  evolution of many binary 
stars. Recently, a WD merger product, \cas,  was identified. \cas\ consists of a central star 
embedded in a 
circular nebula. The analysis of the optical spectrum of the central star revealed that it 
is hot, hydrogen and helium free, and drives an extremely fast wind with a record breaking 
speed. {\new The nebula is visible in infrared and in the  [O\,{\sc iii}]\,$\lambda 5007\,\AA$ 
line images.} No nebula spectroscopy was obtained prior to our observations.
Here we report the first deep X-ray imaging spectroscopic observations of \cas. Both 
the central star and the nebula are detected in X-rays, heralding the WD merger products 
as a new distinct type of strong X-ray sources. Low-resolution X-ray spectra reveal large 
neon, magnesium, silicon, and sulfur enrichment of the central star and  the nebula.
{\nnew We conclude that IRAS 00500+6713 resulted from a
merger of an ONe and a CO WD, which supports earlier suggestion for a
super-Chandrasekhar mass of this object}. X-ray analysis indicates 
that the merger was associated with an episode of carbon burning and possibly accompanied by a SN\,Iax. 
In X-rays, we observe the point source associated with the merger product while the surrounding
diffuse nebula is a supernova remnant. \cas\ will likely terminate its evolution 
with another peculiar Type\,I supernova, where the final core collapse to a neutron star might be 
induced by electron captures.}


  \maketitle
%

\section{Introduction}

White dwarfs (WD) are the degenerate remnants of stars born with intitial mass 
$M_{\rm init}\lsim 10\,M_\odot$. WDs orbiting each other in a binary system emit 
gravitational waves leading  to the gradual orbit shrinking and the eventual merger.
The WD mergers are accompanied by  explosive events, and the outcome of the 
merger depends on the chemical compostions and masses of involved WDs. Likely the most 
common outcome is a supernova (SN) type Ia which completely disrupts the merger  
product \citep{Maoz2014}. However, when a WD involved in a merger descends from an 
intermediate mass star ($M_{\rm init}\approx 8\,..10\,M_\odot$), the merger  
could eventually lead to a creation of a neutron star  \citep[NS,][]{Saio2004}. 
\citet{Schwab2016} showed that a merger of two carbon-oxygen WDs can result in 
a stable, super-Chandrasekhar mass, object. A creation of a super-Chandrasekhar object
could also result from a merger of even more massive WDs  accompanied by   
a peculiar SN, e.g.\ SN\,Iax \citep[e.g.][]{Foley2013,  Kashyap2018}. 
However, while expected to be numerous, neither the remnants of such SNe nor the 
surviving merger products have been firmly identified.

\begin{figure*}[t]
\centering
\includegraphics[width=1.5\columnwidth]{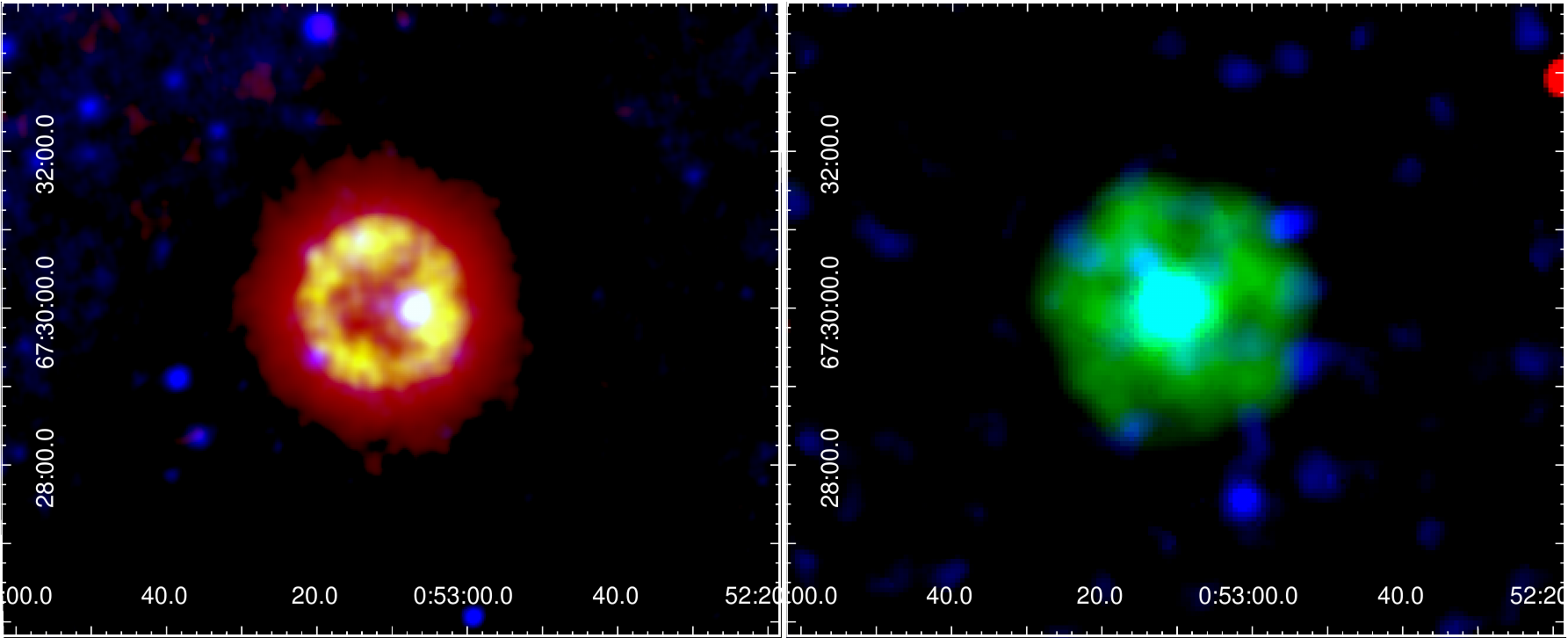}
\caption{Images of \cas\ in mid-IR and X-ray wavelength ranges.  
{\it Left panel:} IR {\it Wide-field Infrared Survey Explorer (WISE)} image: 
red and green correspond to 22\,$\mu$m 
at two intensity scales, blue corresponds to 12\,$\mu$m. A ring-like structure 
with diameter $\sim 2'$ is seen within the nebula;  
{\it Right panel:}  X-ray  EPIC image: 
red: 0.2-0.7\,keV, green: 0.7-1.2\,keV, blue: 1.2-7.0\,keV . The adaptively 
smoothed image shows that X-ray emission uniformly fills the whole $\sim 4'$ 
extent of the IR nebula. At the distance of \cas, $1'$ corresponds to 
$0.86$\,pc. 
The coordinates are in units of RA~(J2000) and Dec.~(J2000) on the horizontal 
and vertical scales, respectively.} 
\label{fig:irximage}
\end{figure*}

{\nnew Recently, \citet{Gv2019} claimed that \cas\ is a super-Chandrasekhar mass object.} \
This object consists of a central star embedded in a circular nebula seen in 
mid-infrared (IR, Fig.\,\ref{fig:irximage}) {\new 
and [O\,{\sc iii}] narrow band filter \citep{Kronberger2014}}. 
{\new The optical spectrum is dominated by strong and broad oxygen emission lines and in this 
respect resembles spectra of WO-type stars. The spectral analysis revealed 
that the central star is hot, hydrogen and helium free, consists 
mainly of carbon and oxygen, and drives a wind with a record breaking speed (Table\,\ref{tab:stpar}).} 
It was suggested that \cas\ resulted from the merger of two CO WDs, although a possibility 
was reserved that a higher mass ONe WD participated in the merger. 
No spectra of the nebula could be obtained and its chemical composition was not known preventing firm 
conclusions on the nature of WDs involved in the merger process and the fate of the merger product.

In this Letter we report first deep X-ray observation of \cas\ and the first spectroscopic 
investigation of the nebula. In section~\ref{sec:obs} we
describe the new X-ray data.  An X-ray spectroscopic analysis of the  
central star is  given in section~\ref{sec:xcs}, while the first nebula spectra
are analyzed in section~\ref{sec:neb}. Section~\ref{sec:conc} presents our explanations of the 
results and the concluding remarks, while detailed description of the X-ray fitting procedure
is presented in the Appendix.

\section{Observations: the central star and its nebula are luminous X-ray sources}
\label{sec:obs}

Observations were obtained 
with the X-Ray Multi-Mirror Mission (\xmm) of the European Space Agency (ESA). 
The three X-ray telescopes of \xmm\
illuminate five different instruments, which always operate simultaneously and 
independently. The useful data were obtained with the three focal instruments: MOS1, 
MOS2 and pn, which together form the European Photon Imaging Camera (EPIC). 
The EPIC instruments have a broad wavelength coverage  of $1.2-60$\,\AA, 
and allow low-resolution spectroscopy with ($E/\Delta E \approx 20-50$). 
The log of the \xmm\ observations is given in Table\,\ref{tab:log}. 
 After rejecting high-background time intervals, the cumulative useful 
exposure time was $\approx 18$\,ks for the EPIC pn and $\approx 31$\,ks 
for the EPIC MOS cameras. The data were analysis using the  \xmm\ data analysis
package SAS\footnote{\url{www.cosmos.esa.int/web/xmm-newton/what-is-sas}}.
Throughout this paper, X-ray fluxes and luminosities are given in the full
$0.2$--$12.0$\,keV energy band,  unless specified otherwise.

The \xmm\ observations
revealed astonishing X-ray properties of \cas\ (Fig.\,\ref{fig:irximage} and 
Table\,\ref{tab:stpar}).
Both the central star and the nebula are clearly detected, heralding the
WD merger products as a new distinct type of strong X-ray sources. 

\section{Analysis of the central star X-ray spectra uncovers carbon and oxygen burning ashes}
\label{sec:xcs}

{\new The central star is more X-ray luminous than single massive OB and 
Wolf-Rayet (WR) stars \citep{Nebot2018}. The optical spectrum of the central star 
is formed in its powerful wind similarly to the massive H- and He-free WO stars, 
yet its X-ray luminosity is orders of magnitude higher compared to the latter \citep{Oskinova2009}.}

\begin{figure*}[]
\centering
\includegraphics[width=0.6\columnwidth, angle=90]{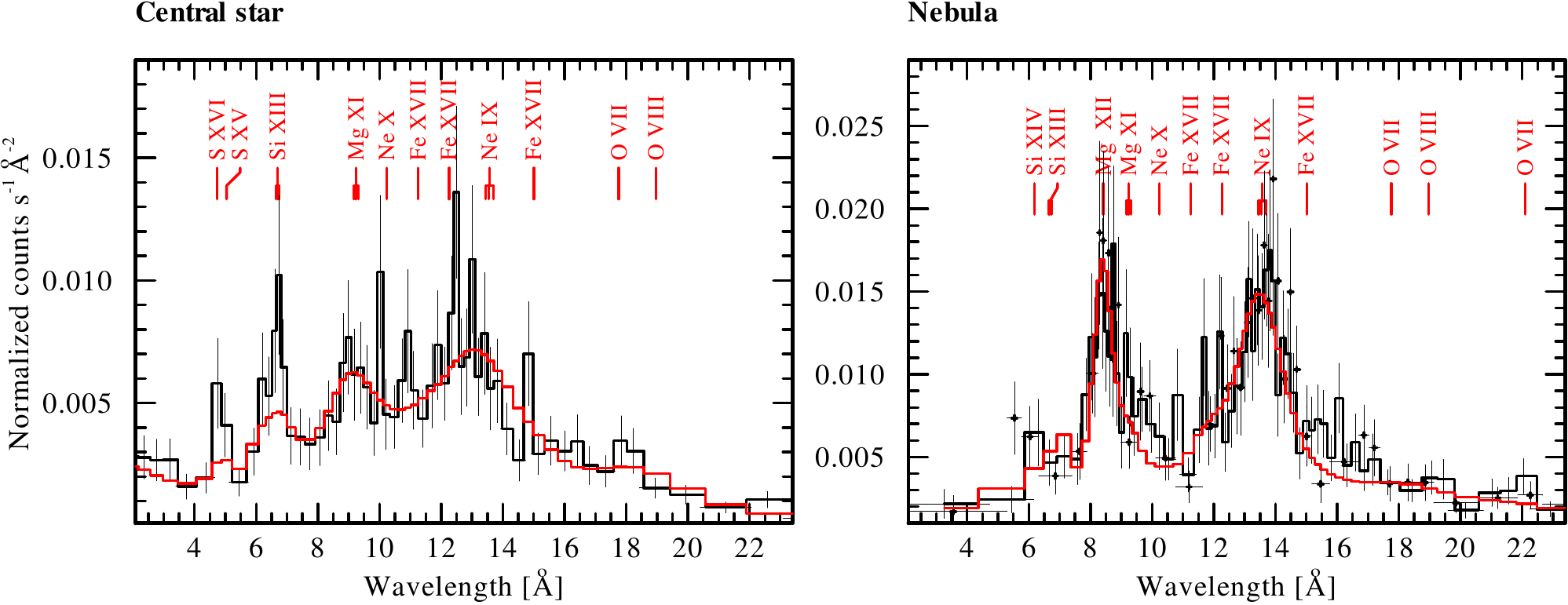}
\caption{Low resolution observed {\new (source plus background) EPIC pn X-ray spectra 
of the central star (left panel) and the nebula (right panel) in \cas. The error bars correspond
to 3$\sigma$. Red histograms show the best-fitting spectral models  
(see Appendix\,\ref{sec:astar} and \ref{sec:aneb} for details on spectral modeling 
and the treatment of the background).  The prominent emission spectral features 
are identified at the wavelengths of strong emission lines  
predicted by the {\em apec} models for plasma with a chemical 
composition and temperatures similar to that of \cas.}} 
\label{fig:specstarneb}
\end{figure*}

The X-ray  spectra (Fig.\,\ref{fig:specstarneb}) and the central star light 
curve (Fig.\,\ref{fig:lc}) were extracted using standard X-ray analysis tools  
(see Appendix) from a 
circle with a diameter $20''$. As the background area, we selected an annulus 
which traces the full extent of the diffuse X-ray emitting nebula around the 
point source. Hence, the contribution of the nebula emission to the spectrum of the 
central star should be small. The pile-up is negligible. 

The pn light curve  binned by 600\,s is shown in Fig.\,\ref{fig:lc}. 
The standard timing analysis procedures were employed to search for a period, but periodicity 
was not detected.

The EPIC spectra were analysed using the 
spectral fitting software {\em Xspec} \citep{Xspec1996}. The spectra 
of collisionally-ionized optically thin plasma were computed with the {\em apec} 
model 
(and its modifications) and the corresponding atomic database  {\em AtomDB} 
\citep{apec2001}. 
 The X-ray spectra are 
dominated by emission lines of Fe, as well as the products of He- and C-burning 
such as 
C, N, O, Mg, Ne, Si, and S (Fig.\,\ref{fig:specstarneb} and Fig.\,\ref{fig:star14}). 
The metal abundances measured from the model fitting to the observed low-resolution 
spectra are shown in Table\,\ref{tab:stpar} using the procedure described in the Appendix\,C. 

\begin{table}
\caption{Parameters of the central star
and the nebula in \cas\ determined from optical \citep{Gv2019} and X-ray (this 
work) spectroscopy. 
The distance 3.1\,kpc is adopted. }
\begin{center}
\begin{tabular}{lr}
\hline \hline
\multicolumn{2}{c}{Central star parameters from optical 
spectroscopy} \\ \hline
$E(B-V)$ [mag]    & $0.84 \pm 0.04$  \\
$T_\ast$ [kK]     & $211^{+40}_{-23}$  \\ 
$\log{L_{\rm bol}/L_\odot}$ & $4.6\pm 0.14$  \\
$\log{(L_{\rm mech}/[{\rm erg\,s}^{-1}}])$  & $\approx 38.4$ \\
Mass-loss rate $\dot{M}$ [$M_\odot$\,yr$^{-1}$] & $(3.5\pm 0.6)\times 10^{-6}$  
\\
Wind velocity $v_\infty$ [km\,s$^{-1}$] & $16\,000\pm 1\,000$  \\ 
$X_{\rm C}$  & $0.2\pm 0.1$  \\
$X_{\rm O}$  & $0.8 \pm 0.1$  \\
$X_{\rm Ne}$ & $0.01$ \\
\hline
\multicolumn{2}{c}{Central star parameters from X-ray 
spectroscopy} \\ \hline
$N_{\rm H}$ [cm$^{-2}$] & $(1.0\pm 0.2)\times 10^{22}$ \\
$L_{\rm X}$ (0.2--12\,keV) [erg\,s$^{-1}$] & $(1.2\pm 0.2) \times 10^{33}$   \\
$\log{L_{\rm X}/L_{\rm bol}}$  & $\approx -5$   \\
$\log{L_{\rm X}/L_{\rm mech}}$ & $\approx -5$   \\
$T_{\rm X}$ [MK] & $1$--$100$   \\ 
$X_{\rm C}$  & $0.15$  \\
$X_{\rm O}$  & $0.61$  \\
$X_{\rm Ne}$ & $0.10\pm 0.03$ \\
$X_{\rm Mg}$ & $0.04 \pm 0.02$ \\
$X_{\rm Si}$ & $0.06 \pm 0.04$ \\ 
$X_{\rm S}$  & $0.04$ \\ \hline
\multicolumn{2}{c}{Nebula parameters from  X-ray spectroscopy} 
\\ \hline
$N_{\rm H}$ [cm$^{-2}$] & $(1.0\pm 0.2)\times 10^{22}$ \\
$L_{\rm X}$ (0.2--12\,keV) [erg\,s$^{-1}$] & $(3.0\pm 0.2) \times 10^{34}$   \\
$T_{\rm X}$ [MK] & $1$--$20$   \\ 
$X_{\rm C}$  & $0.72$ \\
$X_{\rm O}$  & $0.13 \pm 0.06$ \\
$X_{\rm Ne}$ & $0.13 \pm 0.04$  \\
$X_{\rm Mg}$ & $0.02 \pm 0.01$ \\
\hline \hline
\end{tabular}
\end{center}
{\small The O, C, and S abundances were fixed  during the fitting of 
the central star spectra, while only the C abundance was fixed when 
fitting the nebula spectrum (see Appendix).  The error bars correspond to 
$1\sigma$. }

\label{tab:stpar}
\end{table}

X-ray emitting plasma in the central star has a broad 
range of temperatures (Table\,\ref{tab:stpar} and  Table\,\ref{tab:parstarT2}) 
-- assuming purely  thermal plasma,  
temperatures up to $\sim 100$\,MK are required to reproduce the observed 
spectra. 
On the other hand, including a non-thermal
spectral component described by a power-law  improves the spectral fits. In 
this 
case, 
the maximum thermal plasma temperature is $\sim 20$\,MK. Plasma could be heated 
to such 
high temperatures by the shocks in the stellar wind of the central star in \cas,
while the non-thermal radiation could be powered by particle acceleration in 
the 
expected presence of a magnetic field \citep{Gv2019}.

The central star in \cas\ 
has  a high mass-loss rate and a CO-rich wind which should effectively absorb 
X-rays. 
We searched for the presence of K-shell edges in the X-ray spectrum of the 
central star, 
but could not confidently detect them. This rules out an origin of X-ray 
emission
at the base of the wind, or implies that the hot and cool wind components 
are spatially distinct.  

{\new The optical spectrum of the central star was analyzed by means of a non-local 
thermodynamic 
equilibrium stellar atmosphere model (which did not account for X-rays).} The 
carbon 
and oxygen abundances were derived and resulted in the ratio 
(C/C$_\odot$)/(O/O$_\odot$)$\approx  0.6$ (by number) \citep{Gv2019}. 
The optical spectrum analyses also hinted on strong Ne enrichment, 
up to 50\%\ by mass, but magnesium, silicon, and sulfur were not included in the 
 models. In X-ray spectra, the carbon lines  are located redwards 
of 30\,\AA, i.e.\ in the spectral range which suffers from the absorption of 
X-rays 
in the interstellar medium. Consequently, carbon lines are not detected in the 
central 
star X-ray spectra. Hence, during the analysis of X-ray spectra we adopted the 
carbon abundance 
as well as the C/O ratio as derived from optical spectroscopy. 

The spectral models which well reproduce the observed X-ray spectra of the 
central star (Table\,\ref{tab:parstarT2}) 
have similar abundance ratios and require strong enhancement  of carbon-burning 
ashes, as 
well as of Si and S. The emission features corresponding to the blends of the
Si\,{\sc xiii} $\lambda 6.65$\,\AA\ and Si\,{\sc xiv}  $\lambda 6.18$\,\AA\ 
lines, as well as 
of the S\,{\sc xv} $\lambda 5.04$\,\AA\  and S\,{\sc xvi}  $\lambda 4.73$\,\AA\ 
lines are well 
seen in the spectra displayed in 
Fig.\,\ref{fig:specstarneb}. These lines have their peaks at the temperatures 
between 
10\,MK and 26\,MK, i.e.\ in the temperature range  covered by our 
two-temperature spectral 
models. Therefore we believe that the strong Si and S overabundances detected in 
the X-ray spectra
of the  central star are real.  

\section{The lines of carbon burning ashes are detected in the nebula spectra}
\label{sec:neb}
Here we present the first spectroscopy  of the \cas's nebula 
(Fig.\,\ref{fig:specstarneb}).
The nebula images and spectra  were obtained with the help of  
the {\em ESAS} package which also computes the response functions for extended 
sources 
(Appendix B). The extent of the circular X-ray nebula is the same as of the 
IR nebula detected at $22\,\mu{\rm m}$ (Fig.\,\ref{fig:irximage}). \citet{Gv2019} 
attributed the nebula emission to the forbidden lines of [O\,{\sc iv}] $\lambda\,25.89\,\mu$m 
and [Ne\,{\sc v}] $\lambda\lambda\,14.32, 24.32\,\mu$m.  We suggest that the warm 
dust also contributes to the nebula IR emission. The warm dust co-existing with the hot plasma is 
often observed in the supernova remnants (SNR) \citep{Zhou2020}. The X-ray 
nebula is brightest in the medium X-ray band (0.7-1.2\,keV). A number of faint 
X-ray  point sources are superimposed  on the nebula, for now we consider these sources 
as unrelated. 

The X-ray luminosity of the nebula is in the upper range observed from hot bubbles around 
massive WR stars \citep{Toala2017}, and is significantly higher than  the luminosity of diffuse 
gas in planetary nebulae \citep{Chu2001, Kastner2001}.
Furthermore, the plasma temperature significantly exceeds 
the temperatures in WR bubbles or planetary nebulae. X-ray emission of 
these objects  is powered by strong shocks which occur when the  fast wind 
driven by the hot central stars rams into  material of a slow wind ejected at a
previous stellar evolutionary stage. Hydrodynamic simulations of hot WR bubbles and 
planetary nebulae indicate that mixing processes
effectively reduce possible differences in chemical composition of the cool and 
the hot gas components \citep{Volk1985, Kwok2000, Toala2011}. 

The \cas\ nebula spectrum is dominated by two strong emission features corresponding to 
the blends of the Mg\,{\sc xi}\,$\lambda 9.1$\AA\ and Mg\,{\sc xii}\,$\lambda 8.4$\AA\ lines 
and the  Ne\,{\sc ix}\,$\lambda 13.4$\AA\ and Ne\,{\sc x}\,$\lambda 12.1$\AA\ lines 
(Figs.\,\ref{fig:specstarneb}, \ref{fig:neb12}). The blend of the  S\,{\sc xvi}\,$\lambda 4.7$\AA\ 
and S\,{\sc xv}\,$\lambda 5$\AA\ lines is dramatically weaker in the nebula spectrum compared 
to the spectrum of the central star. The emission measure of the hot plasma in the nebula is 
$\sim 10^{55}$\,cm$^{-3}$. Under the crude assumption of a uniform and constant density and the 
nebula composed solely from C, O, Ne and Mg, the total mass of hot gas is  $\sim 0.1\,M_\odot$ 
(see Appendix C).  

\section{\cas: a post WD merger product embedded in a putative SN\,Iax remnant 
and evolving towards an electron capture SN}
\label{sec:conc}
{\new
X-ray spectroscopy allowed us to refine the central star abundances and to determine 
the nebula chemical composition for the first time (Table\,\ref{tab:stpar}). These findings 
call for a reassessment of the \cas\ nature which we consider below.

\subsection{Abundance constrains from X-ray spectroscopy}
The  central star abundances, 
in particular the predominance of Si and S imply a composition resulting 
from incomplete C and O burning. Indeed, assuming that no major 
element remained undetected, Si and S make up $\sim$10\%\ of 
the surface composition, while Ne and Mg constitute up to $\sim$14\%. 
Ne and Mg constitute also up to $\sim$15\%\ of the nebula composition, but 
the S lines are much weaker in the nebula spectrum  (Fig.\,\ref{fig:specstarneb}). 
Furthermore, the C/O and Ne/O ratios appear to be  
different in the central star and the nebula: 
in the latter $X_{\rm C}/X_{\rm O}, X_{\rm Ne}/X_{\rm O} < 1$, 
while in the former $X_{\rm C}/X_{\rm O} \msim 1$, and $X_{\rm Ne}/X_{\rm O} \sim 1$. 

It is important to keep in mind that the wavelength range of X-ray spectra we study
does not include C lines. For the central star,  
C and O abundances were adopted from \citet{Gv2019}. The later work was based on the analyses of optical 
spectra and did not account for X-ray emission (not known at that time). For the nebula,
we calculated a series of spectral models adopting different 
C abundance (two such models are shown in Table\,\ref{tab:parnebT2}), each 
of them resulting in $X_{\rm C}/X_{\rm O} \gg 1$. However, nebula spectra could also be fitted with 
the C and O ratio fixed to be the same as in the central star. 


\subsection{The central star mass, and could it be a partially burned accretor or a donor of 
thermonuclear SNe?}

The surface gravity ($\log{g}$) of the central star cannot be measured  
because no photospheric lines are seen in its optical 
spectrum.  Instead, the \cas\ mass estimate relies on its position 
on the Hertzsprung--Russell diagram (HRD) and the comparison with the evolutionary tracks.  

{\nnew \citet{Gv2019} determined the luminosity of \cas, $\log{L_{\rm bol}/L_\odot}\approx 4.6$, 
from spectral modeling and the {\it Gaia} distance.
Given the H- and He-free composition, and using the WD merger model from \citet{Schwab2016}, 
they concluded that \cas\ is a super-Chandrasekhar object with the mass $\msim 1.5\,M_\odot$.  
We adopt this  mass  estimate.

Taking into account the updated abundances, we suggest that \cas\ was formed by
a merger of a ONe WD and a CO WD with the masses $> 1\,M_\odot$ and $>0.5\,M_\odot$ 
correspondingly. Hence our results corroborate the conclusion on the super-Chandrasekhar 
mass of \cas. The recent evolutionary calculations of WDs with masses up to $1.307\,M_\odot$ show 
that the luminosity of a WD with the ONeMg core $M_{\rm ONeMg}=1.22\,M_\odot$ could approach 
$\log{L_{\rm bol}/L_\odot}\approx 4.5$, however these massive WDs posses H-rich envelopes 
as well as He, while \cas\ is H- and He-free \citep{Lauffer2018}. }

It is also informative to compare \cas\ with other sub-Chandrasekhar mass WDs which are H- and He-free 
and are rich in the ashes of C-, O-, and Si-burning 
\citep[][and ref. therein]{Shen2018, Gaensicke2020MNRAS.496.4079G}.
These runaway WDs (often called LP\,40-365-like and  D$^6$-type) belong to a heterogeneous 
group of objects consisting of  partially burned accretors and puffed-up donors of 
thermonuclear SNe.
The central star in \cas\ does not resemble these objects. 
Without correction for reddening \cas\ has  $M_{\rm G}\approx 2.76$\,mag from the
{\em Gaia}\,DR2 data.
This is significantly smaller than the $M_{\rm G}$ of the runaway D$^6$-type WDs 
discussed by \citet{Shen2018}.  Furthermore, the optical spectra of \cas\ and the 
D$^6$ \& LP\,40-365-like WDs are very different -- while the later show photospheric 
spectra and no trace of winds, the former spectrum is formed in a strong outflow. 
{\nnew  One may speculate that if \cas\ would not collapse in the course of its evolution 
but instead cool, then it could become a WD with an extreme surface composition; an interesting 
question is whether such an object would resemble the LP\,40-365-like and  D$^6$-type SNe survivors. }

\subsection{\cas\ stellar wind, remarks on its possible magnetic field, and ruling 
out a recent SN }

The strong stellar wind in \cas\ corroborates the super-Chandrasekhar nature of its 
central star because objects close to their Eddington limit 
are expected to launch powerful winds \citep{Graefener2013, Sander2020}. 
Furthermore,  wind acceleration could be aided by magnetic field as expected in
 WD mergers \citep[e.g.][]{Beloborodov2014MNRAS.438..169}.
{\nnew It should be noted, that whereas mergers may produce magnetic WDs, there is no 
strong evidence that they always do so. Conversely, there are strongly magnetic WDs in 
binaries \citep{Pala2020}.} 
 
Whether \cas\ is indeed magnetic is not known. However, \citet{Gv2019} invoked the theory of 
rotating magnetic winds \citep{Poe1989} to qualitatively explain the velocity  of the wind 
in \cas, while \citet{Kashiyama2019} used a magneto-hydrodynamic model and demonstrated that 
the optically thick outflow could be launched from the C-burning shell on an ONe core and 
accelerated by the rotating magnetic field. 

The mechanism of the central star X-ray emission is not clear but the very 
high wind velocity provides an ample energy reservoir for plasma heating. The 
detection of a non-thermal component in the X-ray spectrum hints that 
magnetism may also play a role. However, X-rays originating deep in stellar wind 
would hamper wind acceleration and be strongly absorbed -- this is not observed. 
Therefore, we tentatively suggest that X-rays 
are produced in the outer wind regions.  

The wind speed, $16\,000$\,km\,s$^{-1}$, is comparable to 
expansion velocities of SN ejecta. Could it be that what we interpret as a wind is in fact
SNR material coasting at that velocity? To investigate this question, we searched historical 
astronomical data.  
Photographic plate observations obtained in the 
{\em Hamburger Sternwarte} and digitized by the {\em APPLAUSE} 
project\footnote{\url{https://www.plate-archive.org/objects/dr.3/plates/101_1702} \\
\url{https://www.plate-archive.org/objects/dr.3/plates/101_1696}} show 
that  \cas\ had V$\approx$15\,mag in November 
1926, i.e.\ similar to its present day V=15.44\,mag (The Guide Star Catalog 2.3, 
2006) or V=15.23\,mag (The NOMAD-1 Catalog, 2005). The photometric monitoring 
of \cas\ in 2017--2019 shows only small variability around $V=15.49$\,mag, not exceeding 
0.05\,mag (A.\,S.\,Moskvitin, private communication). Thus, the 
brightness in V remained stable within 0.5\,mag during 
100\,yr. Moreover, the optical spectra do not show any evolution during 
the last three years \citep[2017--2020,][]{Gv2019, 2020arXiv200912380G}. Hence
the central star is a stable object and its strong stellar wind is not 
a sporadic  ejecta.  

\subsection{The high carbon abundance in \cas\ argues against its nature as an 
electron capture SN}

Among a small group of known H- and He-free objects are the stellar remnants of  
electron-capture SNe (ECSNe).
\citet{Jones2019} studied O-Ne deflagrations in WDs as an underlying mechanism of 
ECSNe. Their 3D hydrodynamic models and the post-processing predicting the ejecta 
composition imply that ECSNe (including accretion-induced collapse (AIC) of ONe WDs), 
could be partial thermonuclear explosions leaving behind a bound ONeFe WD.  
However,  this model predicts a very low C abundance 
(see e.g. their Table\,7). If future analyses of the \cas\ UV spectra will 
reveal significantly lower C abundance than $X_{\rm C}=0.2\pm 0.1$ derived by \citet{Gv2019},
and if it would be shown that ONeFe WDs could have luminosities as high as\cas, then a bound  
remnant of an ECSN and its SNR could become an interesting possibility to explain \cas. 
However, based on the current data and models this channel appears unlikely.     

\subsection{SNe type Iax in a single-degenerate scenario}

SNe\,Iax are a subset of peculiar SNe\,Ia, which have low luminosities, and low ejecta 
velocities and masses \citep{Foley2013}. In their review of SN\,Iax, \citet{Jha2017} concluded
that they results from a CO or a hybrid CONe WD that accretes from 
a He-star companion, approaches the Chandrasekhar mass, and explodes as a deflagration 
that may leave a bound remnant. \citet{Fink2014} carried out a study of a CO WD  
explosion  in a single-degenerate scenario. The major result of their 3D hydrodynamic 
models is the occurrence of a bound remnant mostly comprised of the unburnt matter and the
ejection of the hot
deflagration ashes at velocities up to 14\,000\,km\,s$^{-1}$. Unburnt CO material and $^{56}$Ni  
 can be found at all ejecta velocities. Furthermore, the model predicts that 
the outer layers of the bound remnants are enriched with the iron group  and 
intermediate mass (Si, S) elements. In the ejecta, the abundances  of Ne and Mg 
are significantly lower than O, while O/C$\sim 1$.  

\citet{Leung2020} carried out 2D simulations of the propagation of 
deflagration which leaves a small WD remnant behind and eject nucleosynthesis materials. 
The nucleosynthesis and explosion energy depend on the central densities 
and compositions of the WDs, and on the flame prescriptions. The model predicts a low mass 
WD remnants similar to LP\,40-365-like and related 
objects. Furthermore, \citet{Leung2020} considered massive ONeMg WDs resulting from
a super-AGB star evolution  when the core or 
shell O-burning is ignited by electron capture and can trigger oxygen deflagration. 
In these models ejecta have $X_{\rm Ne}/X_{\rm O}\lsim 1$ which is very promising 
when comparing with the \cas\ nebula properties. 

Nevertheless, a SN\,Iax in a single degenerate scenario does not seem to be a likely  
explanation for \cas. {\em Firstly}, a He-star donor is not observed; 
{\em secondly,} \cas\ did not experienced a SN in the last 100\,yr;  
{\em thirdly,} the central star is not a low mass WD.  
 
\subsection{\cas\ as a result of ONe and CO WD merger accompanied by a SN\,Iax and evolving
towards an ECSN}
\label{sec:one}

In agreement with \citet{Gv2019} suggestion, we favor a double-degenerate channel. 
The hydrodynamic models of a merging  ONe ($1.2\,M_\odot$) and CO ($0.6\,M_\odot$)  WD binary 
\citep{Loren2009} show a mild nuclear processing of material from the CO dwarf, with 
some Ne and Mg production, but not compatible with the high Si and S abundances in \cas. On the 
other hand, the model remnant WD mass of $1.5\,M_\odot$ fits well to the mass of \cas\
as estimated from its current luminosity.

The higher mass ONe+CO WD merger models \citep{Kashyap2018}
predict that the less massive but larger CO WD
is tidally disrupted and forms a hot, low-density accretion disk
around the ONe WD. The ignition and explosive disruption of
this disk produces a low-energy ($\sim 10^{49}$\,erg) SN, putatively of
Type\,Iax, which leaves the ONe WD largely intact.
In these models, the remaining WD re-accretes part of the explosive ejecta,
which are highly enriched in Si and S (their
Figs. 5 and 6), together with Ne, Mg and unprocessed C and O.
The final mass of the star left behind in this model is $\sim 
2.2\,M_\odot$ and the predicted $X_{\rm Ne}/X_{\rm O}\approx  0.04$
(see their Table\,1), i.e.\ lower than we find from X-ray spectroscopy. 

In the ONe and CO WD merger accompanied by a SN\,Iax scenario, the \cas\ nebula is 
 a SNR. Scaling relations based on the Sedov solution \citep{Bor2001, Osk2005} 
yield the SNR age of $\sim 1000$\,yr.  
SNe\,Iax  can be as dim as $M_{\rm V}=-14\,$mag. The SN would then only appear as bright 
as Sirius  ($V=-1.5\,$mag), and with a duration of just 2 weeks it could have been  easily
missed in the last millennium.

The inner ring-like shell seen in the IR image (Fig.\,1, left panel) with the 
radius $R_{\rm sh}\approx 1$\,pc could have been  created by the current fast wind 
of the central star. The  distance at which the wind 
ram pressure is balanced by the thermal pressure in the SNR:
$
R_0 =\sqrt{{\dot{M} v_\infty / 4\pi P_{\rm th}}} \, , 
\label{eqn:bow}
$
where $P_{\rm th}=(\gamma -1)E_{\rm th}/V$, $\gamma=5/3$, $E_{\rm th}$ is the thermal energy of 
the SN blast wave, $V=(4\pi/3)R_{\rm SNR} ^3$, and $R_{\rm SNR}$ is the radius of the nebula.
In the Sedov phase about 72\%\ of the kinetic energy produced by the SN 
explosion, $E_0$, is converted into $E_{\rm th}$.
Adopting $E_0=2.2\times10^{49}$\,erg \citep[as in ][]{Kashyap2018},  $E_{\rm th}=1.6\times10^{49}$\,erg.
Using $\dot{M}$ and $v_\infty$ from Table\,\ref{tab:stpar}, and $R_{\rm SNR}=1.6$\,pc, 
one finds that 
$P_{\rm th}=2\times10^{-8} \, {\rm dyne} \, {\rm cm}^2$ and 
$R_0=0.4$ pc. This is 4 times smaller than the observed
$R_{\rm sh}$. This could be explained if $E_0$ was smaller or $P_{\rm th}$ is not a 
constant value but grows with radius.   
For $E_0=2.2\times10^{49}$\,erg and $n_{\rm ISM} = 1 \, {\rm cm}^{-3}$, the age 
of the SNR is only $t_{\rm SNR}\approx 350 \, {\rm yr}$ , while for a 
factor of ten lower value of $E_0$, the SNR age is $t_{\rm SNR}\approx1100 \, {\rm yr}$. 
These should be the characteristic times for the merger product to contract to its current size 
and to develop its very fast wind.

\citet{Wang2020RAA....20..135W}  provided estimates of the Galactic rates of
ONe+CO WD mergers. According to their predictions, there should 
be $10^6$--$10^7$ ONe+CO WD binaries in the galaxy. Interestingly, such WD binaries belong 
to relatively young stellar population (50--100\,Myr old) in agreement with the host 
populations of SNe\,Iax \citep{Jha2017}.

Clearly, more WD merger models are needed to pin down the progenitor masses of 
the two WDs which produced \cas\ and explain its abundances. 
Whereas current models of WD mergers do not include stellar
winds, the discovery of an extreme WO-type wind from the central star in \cas\
demonstrates the urgent need to do so, at least in models of post WD-merger 
evolution.

The farther evolution of \cas\ is spectacular. 
Based on its empirical mass loss rate and the expected short remaining life time 
of several $1000\,$yr, the mass of \cas\ will likely remain above the Chandrasekhar limit.
Its fate will therefore be to undergo core collapse and to form a neutron star. 
In the course of this event, \cas\ will manage to produce its second SN, possibly 
in the form of a fast blue optical transient  \citep{Dessart2006, Lyutikov2019}.

\begin{acknowledgements} {\new The authors are grateful for the very useful and insightful 
referee report which allowed to significantly improve the paper. We thank J.M. Drudis for 
bringing to our attention the paper by \citet{Kronberger2014}.}
L.M.O. acknowledges financial support by the Deutsches Zentrum 
f\"ur Luft und Raumfahrt (DLR) grant FKZ 50 OR 1809, and partial support by the Russian
Government Program of Competitive Growth of Kazan Federal
University. V.V.G. acknowledges support from the Russian Science Foundation under grant 19-12-00383. 
G.G. thanks the Deutsch Forschungsgemeinschaft for financial support
under Grant No.\ GR\,1717/5-1. {\new Funding for APPLAUSE has been provided by 
DFG, AIP, Dr. Remeis Sternwarte Bamberg, the Hamburger Sternwarte and Tartu Observatory. 
Plate material also has been made available from Th\"uringer Landessternwarte Tautenburg.}
\end{acknowledgements}

\appendix

\section{X-ray properties of the  central star}
\label{sec:astar}

\begin{table}
\begin{center}
\caption[ ]{Log of the \xmm\  observations  of \cas}
\begin{tabular}[]{ccc}
\hline
\hline
ObsID & Start-date & exp. time [ks] \\
\hline 
0841640101 & 2019-07-08  & 19.7 \\
0841640201 &2019-07-24  & 16.8  \\
\hline
\hline
\end{tabular}
\label{tab:log}
\end{center}
\end{table}

\begin{figure}[t]
\centering
\includegraphics[width=0.7\columnwidth]{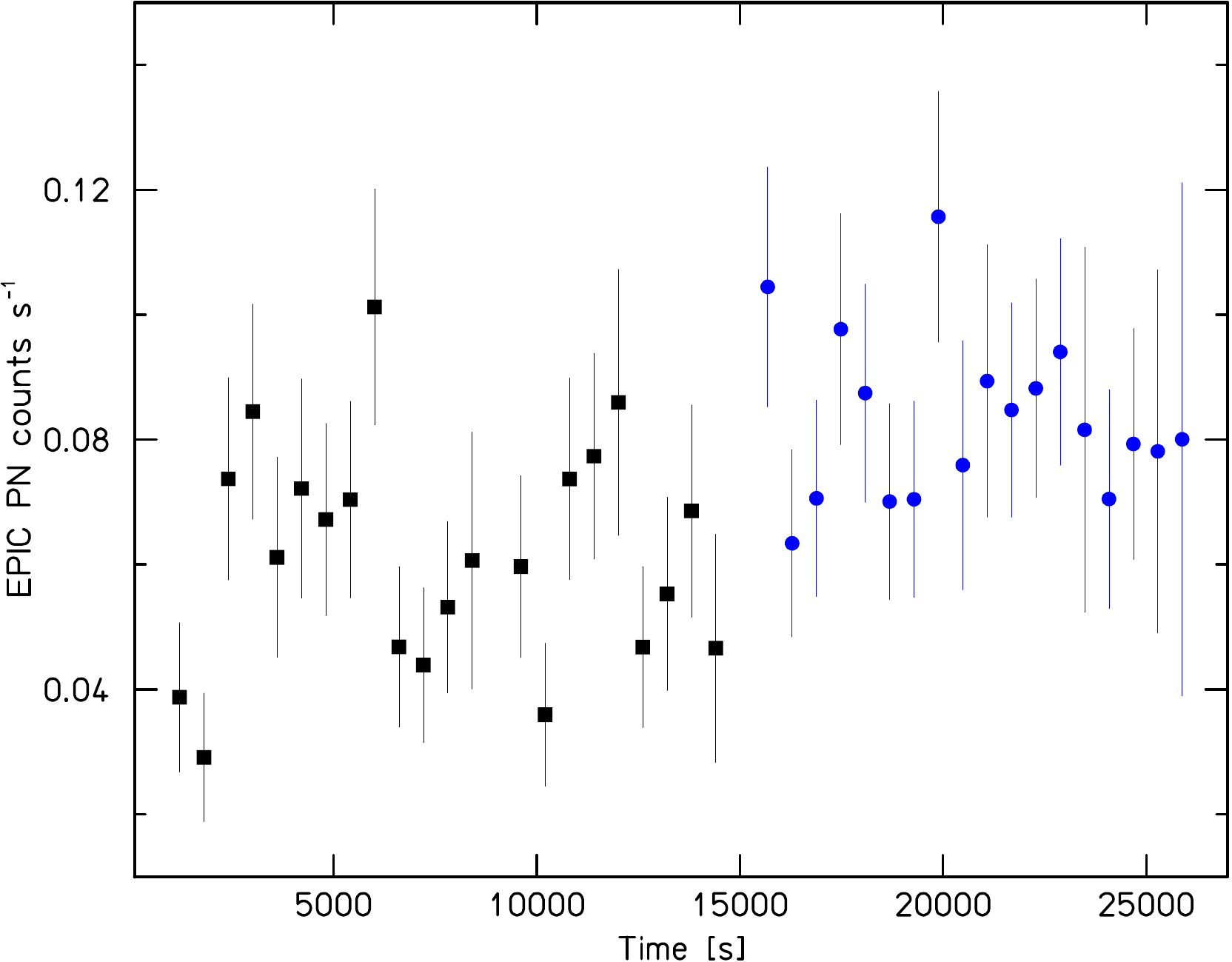}
\caption{\xmm\ EPIC pn  light curve of the central star in \cas.
The light-curve is background corrected and is binned by 600\,s, the  
error bars correspond to 3$\sigma$.  The Y-axis is the time after the 
start of the observation. Black and blue data points refer to the first and 
second observation respectively (see Table\,\ref{tab:log}). For the second 
observation, the X-axis is shifted by 15\,000\,s for clarity.  } 
\label{fig:lc}
\end{figure}

\begin{figure}[h]
\centering
\includegraphics[width=0.95\columnwidth]{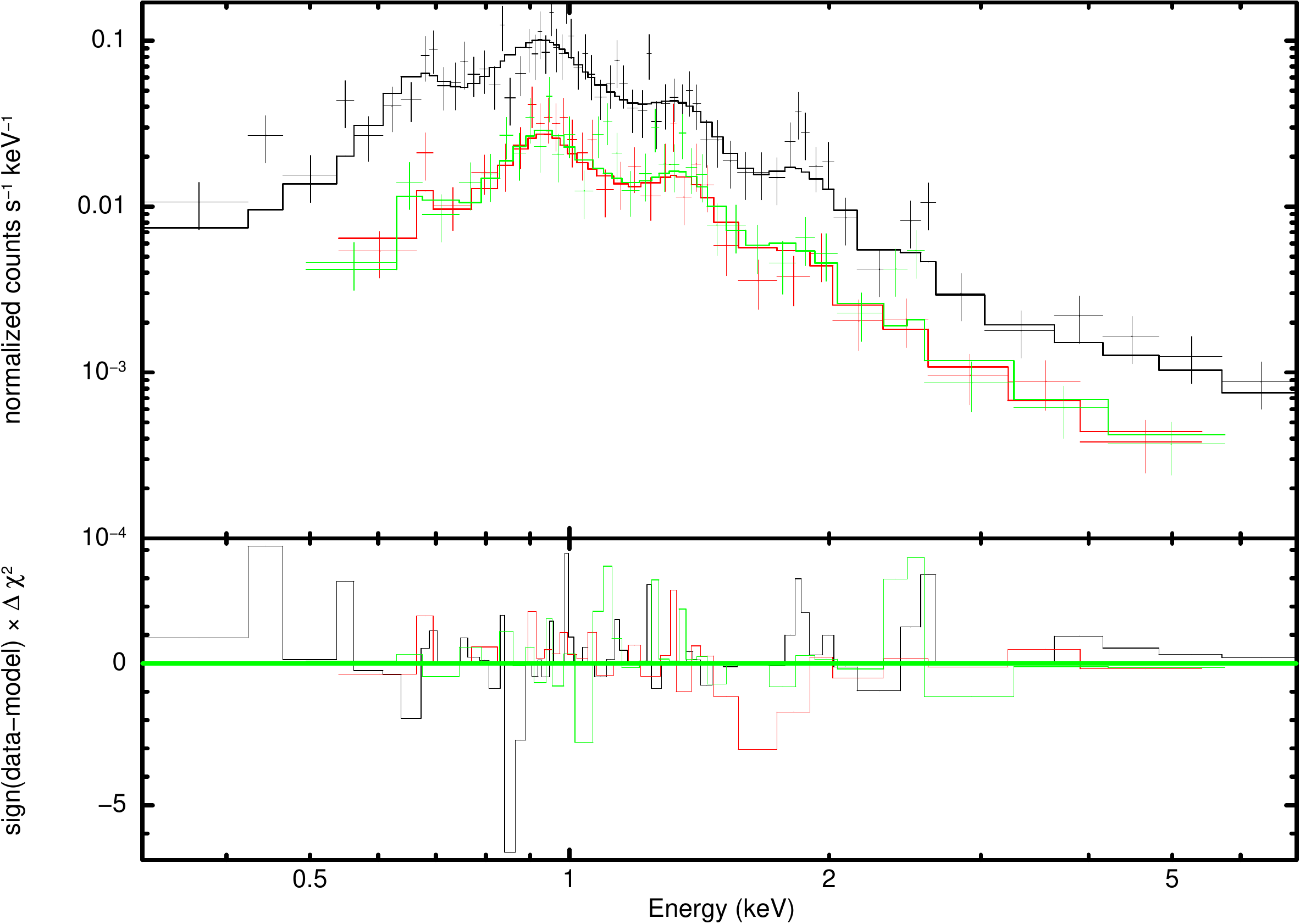}
\caption{Low-resolution X-ray spectra of the central star in \cas. 
The EPIC pn (black data points), MOS1 (red data points) and MOS2 (green data points)
spectra merged over the full exposure time are displayed. The black, red
and green histograms show the best fitting model of a two-temperature
plasma and a non-thermal components, and the residuals as signed contributions.
} 
\label{fig:star14}
\end{figure}
{
X-ray spectra of the central star were extracted from a circle with $r=10''$ 
centered on the coordinates of \cas. As the background area, we selected the annulus 
around the point source which traces the full extent of the X-ray nebula with the outer radius 
$40''$. Hence, the contribution of nebula emission to the spectrum of the central star should 
be small or negligible. The event lists were filtered to exclude the intervals of flaring 
particle background exceeding 0.4\,s$^{-1}$ for pn, and 0.35\,s$^{-1}$ for the MOS cameras.
The SAS task {\em evselect} was used to extract the spectra in the 0.2-10.0\,keV range 
and setting the binning factor (via the parameter {\em spectralbinsize}) to 5.} 
The spectra obtained during each of the two  observations (Table\,\ref{tab:log}) were combined 
using the {\it epicspeccombine} SAS task. The three resulting spectra are shown in  
Fig.\,\ref{fig:star14} were fitted using the {\em apec}, where the element abundances are 
normalized to solar values relative to hydrogen. 
Having this caveat in mind, the metal abundances measured from the model fitting to the observed 
low-resolution spectra shown in Tables\,\ref{tab:parstarT2} and \ref{tab:parnebT2} are relative to solar 
values \cite{Asplund2009}.

Statistically acceptable fits to the observed spectra of the central star are obtained with 
the absorbed \citep[by using the {\it tbabs} model,][]{Wilms2000}  
multi-temperature {\em bvtapec}\footnote{A velocity- and thermally-broadened 
emission spectrum from collisionally-ionized diffuse gas calculated from the AtomDB atomic
database, where we assumed identical ``continuum'' and ``line'' temperatures.}  spectral model. 
The stellar wind of \cas\ is exceptionally fast, with 
$v_\infty=16000$\,km\,s$^{-1}$. If the hot plasma is expanding with similar velocity then 
emission lines observed in the X-ray spectrum are broad. The spectral resolution of EPIC cameras 
($\Delta E = 20-50$)  is not sufficient to resolve even such broad lines. Nevertheless, we found 
that accounting for line broadening (by setting the velocity parameter 
to $16000$\,km\,s$^{-1}$) improves the fitting statistics -- { for 124 d.o.f., the reduced $\chi^2$=0.91 
without including the line broadening, while it drops to $\chi^2$=0.89 when the line broadening is accounted 
for}. 

\begin{table}
\begin{center}
\caption[ ]{
X-ray spectral properties of the central star in \cas\ from fitting its low-resolution EPIC spectra.
The ion and the continuum temperatures are assumed to be equal. The line broadening is set to 
$v=16000$\,km\,s$^{-1}$. The abundances shown 
without errors were not fitted but frozen during the fitting process. The abundances 
which are not shown in the table are at their solar values, expect He and N which are vanishingly 
small. Observed flux and de-reddened luminosity are in 0.2-12.0\,keV band assuming $d=3.1$\,kpc.}
\label{tab:star}
\begin{tabular}[]{lcc}
\hline
\hline
Model parameter & \multicolumn{2}{c}{fit value}   \\
\hline
spectral model & {\small\it 2Tbvtapec} &  {\small\it 2Tbvtapec+power} \\ \hline
$N_{\mathrm H}$ [$10^{22}$ cm$^{-2}$]  & $1.0\pm 0.2$    & $1.0 \pm 0.2$  \\
$kT_1$ [keV]                           & $0.27 \pm 0.04$ & $0.27 \pm 0.04$\\
$EM_1$ [$10^{53}$ cm$^{-3}$]           & $2 \pm 1   $    & $2  \pm  1$    \\
$kT_2$ [keV]                           & $6 \pm 2$       & $1.6 \pm 1.1$  \\
$EM_2$ [$10^{53}$ cm$^{-3}$]           & $0.8 \pm 0.1$   & $0.2 \pm  0.1$ \\
C  & $600$ & $600$   \\        
O  & $1000$  & $1000$ \\
Ne & $800 \pm 200$ & $750 \pm 200$  \\
Mg & $600 \pm 300$ & $600 \pm 250$  \\
Si & $1000 \pm 900$ & $800 \pm 520$  \\
S  & $1000$        & $1000$        \\ 
$\alpha$   &           & 0.9   \\
K (ph\,keV$^{-1}$\,cm$^{-2}$\,s$^{-1}$) & & $(4\pm 1)\times 10^{-6}$  \\  
reduced $\chi^2$ for 124 d.o.f. &  $0.93$ & $0.89$  \\ \hline
Model $F_{\rm X}$ [\flux] & \multicolumn{2}{c}{$(1.7\pm 0.2)\times 10^{-13}$} \\
Luminosity $L_{\rm X}$ [erg\,s$^{-1}$] & \multicolumn{2}{c}{$(1.2 \pm 0.2) \times 10^{33}$}  \\
$\log{L_{\rm X}/L_{\rm bol}}$  & \multicolumn{2}{c}{$-5$}   \\
\hline
\hline
\end{tabular}
\label{tab:parstarT2}
\end{center}
\end{table}

To determine the abundances, we first tested a model where the interstellar absorbing column density, 
$N_{\rm H}$, and the C, O, N, Mg abundances were considered as free parameters; however no meaningful 
constraints on these parameters could be derived. The reddening to the star, $E(B-V)$, is known from the 
optical (Table\,\ref{tab:stpar}). Hence, we tested a model with the 
fixed $N_{\rm H}=5.8 \times 10^{21}E(B-V)$ \citep{Bohlin1978}, however the metal abundances still 
could not be constrained. 
The C\,{\sc v}\,$\lambda 41$\,\AA\ and C\,{\sc vi} Ly$\alpha$ $\lambda 33$\,\AA\ lines are 
located in the part of the \xmm\ spectrum which has a low signal-to-noise ratio and which 
suffers from the absorption. Hence, X-ray spectral models are not sensitive to carbon abundances. 
Therefore, we decided to reduce the parameter space by using the abundance measurements from optical 
spectroscopy with the carbon and oxygen abundances determined as 
C/C$_\odot \approx 930$ and O/O$_\odot \approx 1500$ \citep[by number,][]{Gv2019}. Unfortunately, 
in {\it Xspec} the 
maximum value of an element abundance { can not exceed} $1000$. Therefore, we froze the 
O abundance to $1000$ and set the C abundance to $600$ to preserve the C/O ratio derived from
the optical spectra. 

\begin{figure}[h]
\centering
\includegraphics[width=0.95\columnwidth]{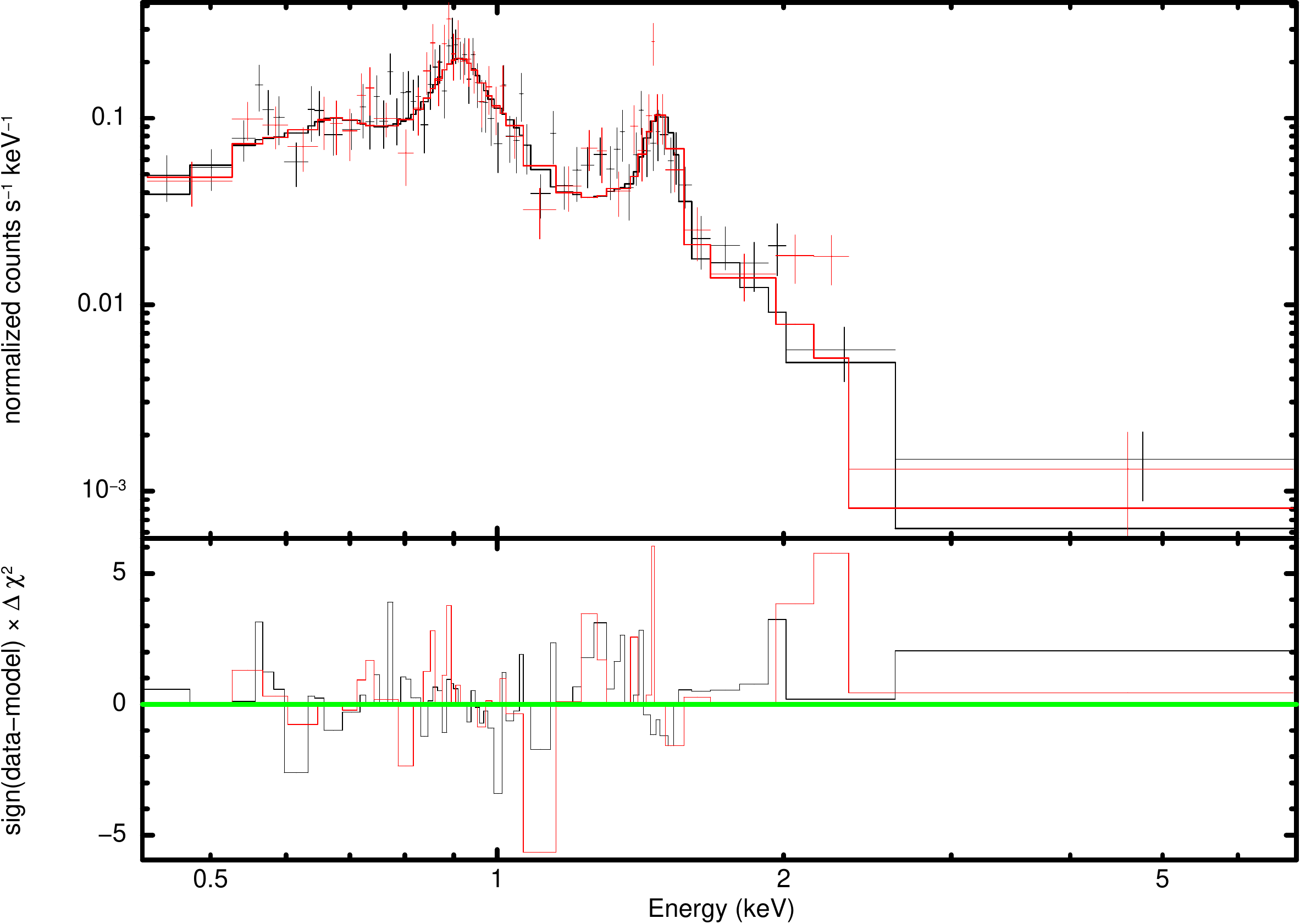}
\caption{Low-resolution X-ray spectra of the nebula in \cas. 
The EPIC pn (black and red data points) 
spectra obtained during two pointings are displayed. The histograms show the 
best-fitting model (second column in Table\,\ref{tab:parstarT2})  of a two-temperature plasma, 
and the residuals as signed contributions.
} 
\label{fig:neb12}
\end{figure}

A two temperature model with Ne, Mg, Si, and S abundances being fitting parameters 
tied among both temperature components produces an excellent quality fit, indicating that an
S abundance is very high but very poorly constrained. The strong S line blend at 
$\lambda\lambda 4.7, 5.0$\,\AA\ is well seen by eye. To roughly match the line strengths, 
we freeze the S abundance to the maximum possible value S/S$_\odot$=1000. This immediately improves the 
fitting statistics. The S\,{\sc xvi} Ly$\alpha$ has the maximum emissivity at the temperature 
25\,MK.  This temperature regime is covered by the hottest plasma component in our models, hence we 
believe that the large S abundance deduced from spectral fitting is real.
The fitting statistics improve further when a non-thermal spectral energy component in included in 
the model (Table\,\ref{tab:parstarT2} and Fig.\,\ref{fig:star14}).
Finally, the spectral models which include the K-shell edges were considered but no meaningful 
constraints on the presence of the edges could be obtained. 

\begin{figure}[h]
\centering
\includegraphics[width=0.95\columnwidth]{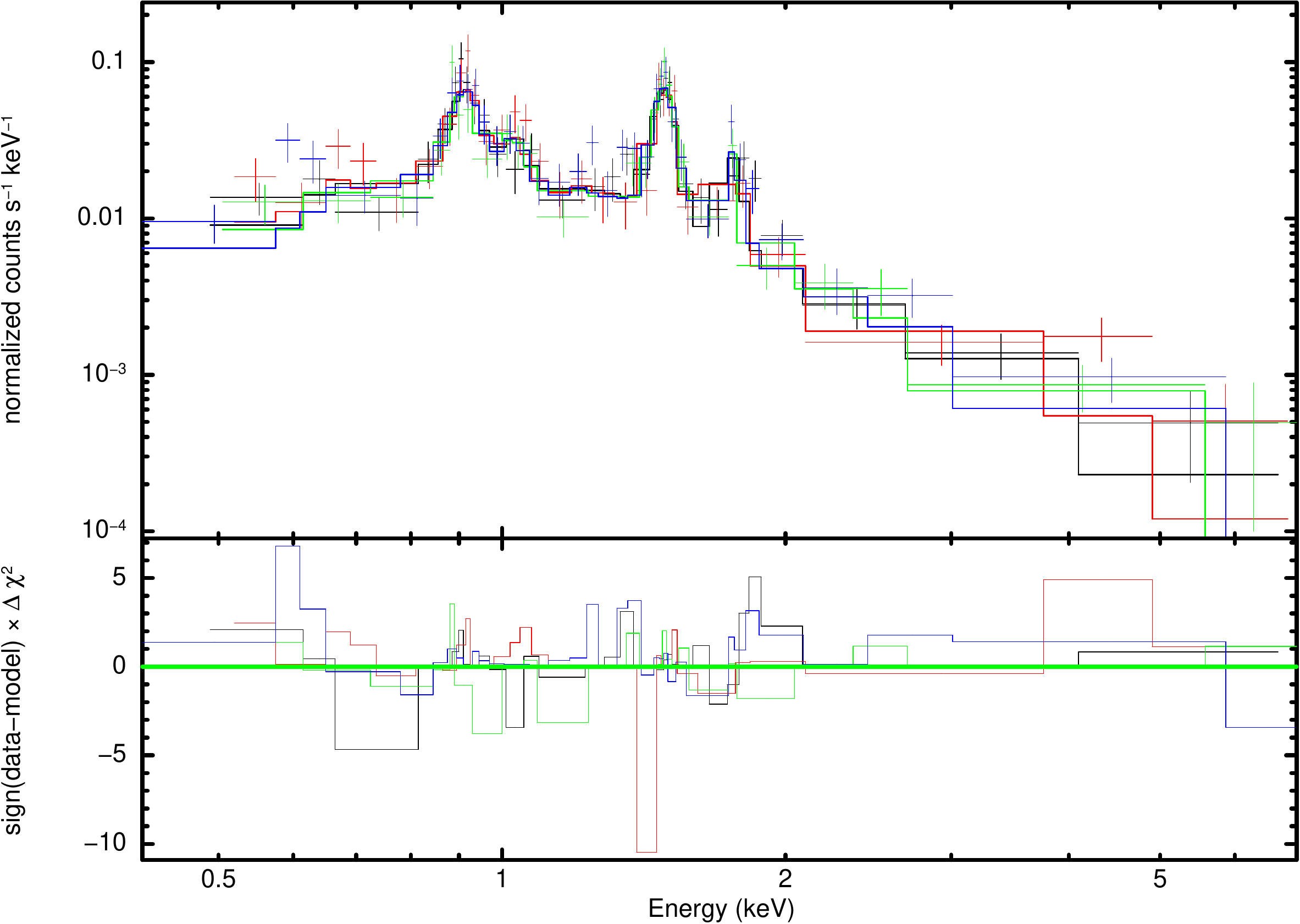}
\caption{The same as in Fig.\,\ref{fig:neb12} but for EPIC MOS1 and MOS2
spectra obtained during two pointings are displayed. The histograms show the 
best-fitting model (fourth column in Table\,\ref{tab:parstarT2} of a two-temperature plasma, 
and the residuals as signed contributions.
} 
\label{fig:nebmos}
\end{figure}

\section{X-ray properties of the nebula}
\label{sec:aneb}

The Extended Source Analysis Software package (ESAS) \cite{Snowden2004, Snowden2008} which is 
integrated in the SAS was employed  to obtain the images and spectra of the diffuse emission.
Following the data reductions steps prescribed by the \xmm\ data analysis threads and the ESAS 
cookbook\footnote{\url{https://xmm-tools.cosmos.esa.int/external/sas/current/doc/esas/}}, 
the EPIC images were created in ``soft'' (0.2--0.7\,keV), ``medium'' (0.7--1.2\,keV), and
``hard'' (1.2--7.0\,keV) energy bands.  The background was modelled and subtracted, and 
the individual images were  merged after  correcting them by their corresponding  exposure maps. 
Each image in each band was adaptively smoothed requesting 50 counts under the smoothing kernel. 
The resultant exposure-corrected and background-subtracted colour-composite image of the sky in the 
vicinity of \cas\ is shown in Fig.\,\ref{fig:irximage}. 

The background and the point-source corrected spectra of the nebula, as well as the corresponding 
response matrices, were extracted for each camera (MOS1, MOS2, pn) and for each observation 
(Fig.\,\ref{fig:neb12}) using the {\it pn-spectra}, {\it pn-back}, and {\it mos-spectra}, {\it mos-back}
tasks in the ESAS package.  All spectra have been fitted simultaneously with the abundances tied 
between different model temperature components. 
{\new  As recommended by the ESAS Cookbook, the spectra were fitted in the 0.4--7.0\,keV range.  
To account  for the instrumental background produced by the fluorescence lines of Al K$\alpha$ 
(E$\sim1.49$\,keV) and Si K$\alpha$ (E$\sim1.75$\,keV) lines in the MOS cameras and  the Al K$\alpha$ 
in the pn camera we used the EPIC instrumental background files produced with the filter wheel 
equipped in the ``CLOSED''  position. For this purpose we employed the SAS task {\em  evqpd} and 
produced a tailored  ``Filter Wheel Closed'' (FWC) event file
for both our observations.  Spectra were extracted from the FWC files in the same area and at the 
same detector position as
our science nebula spectra. The Al\,K$\alpha$ and Si\,K$alpha$ lines were fitted as Gaussians and 
thus   their 
normalizations were determined. As a next steps, these lines were added to the spectral model. 
As a result,  the model {\it tbabs(vapec+vapec)+gauss(1.49\,keV)}  was used to fit pn spectra 
(Fig.\,\ref{fig:neb12}, while 
 {\it tbabs(vapec+vapec)+gauss(1.49\,keV)+gauss(1.75\,keV)}  was used for the MOS spectra 
 (Fig.\,\ref{fig:nebmos}). }

\begin{table*}
\begin{center}
\caption[ ]{
\new X-ray spectral properties of the nebula in \cas\ from fitting its low-resolution EPIC pn 
(Fig.\,\ref{fig:neb12}) and MOS (Fig.\,\ref{fig:nebmos}) spectra . The spectral model is a 
the two temperature collisional plasma  
corrected for the interstellar (ISM) absorption {\it tbabs(vapec+vapec)}). The Gaussian lines describing  
Al K$\alpha$ and Si K$\alpha$ lines for the MOS and the Al K$\alpha$ for the pn cameras 
are explicitly added to the background (see text). 
{\em First column}:  the model parameters  from fitting pn spectra with a carbon abundance frozen to  C=100.  
{\em Second column}:  the same as in the first column model but with a carbon abundance frozen to  C=600.
{\em Third column}:  the same as in the second column model but for EPIC MOS.
The abundances which are not shown in the table were kept at their solar values. 
Observed flux and de-reddened luminosity are in the 0.2-12.0\,keV band assuming $d=3.1$\,kpc.}
\label{tab:star}
\begin{tabular}[]{lccc}
\hline
\hline
Model parameter & \multicolumn{3}{c}{fit value}   \\
\hline 
\hline
$N_{\mathrm H}$ [$10^{22}$ cm$^{-2}$]  & $1.1 \pm 0.3$  & $1.1 \pm 0.3$  &  $1$  \\
$kT_1$ [keV]                  & $0.11\pm 0.0$   & $0.11 \pm 0.01$          & $0.13\pm 0.01$ \\
$EM_1$ [$10^{54}$ cm$^{-3}$]  & $260\pm 90$     & $41 \pm 16$ & $27 \pm 6$   \\
$kT_2$ [keV]                  & $0.8 \pm 0.1$   & $1.1 \pm 0.2$  & $1.60  \pm 0.15$ \\
$EM_2$ [$10^{54}$ cm$^{-3}$]  & $3.2 \pm 0.7$   & $0.9\pm  0.2$  & $1.6 \pm 0.1$    \\
C  & $100$ & $600$   & $600$ \\        
O  & $9\pm 4$    & $46\pm 21$  & $46$  \\
Ne & $37 \pm 13$ & $201 \pm 70$  &  $330\pm 100$ \\
Mg & $15 \pm 5 $ & $69  \pm 26$   & $33\pm 19$  \\
reduced $\chi^2$   &  $1.0$ (232 d.o.f.)  & $1.0$ (232 d.o.f.) & $1.0$ (189 d.o.f.) \\  \hline
Model $F_{\rm X}$ [\flux] & \multicolumn{2}{c}{$ (1.8  \pm 0.1) \times 10^{-13}$}  &  $ (2.2  \pm 0.1) \times 10^{-13}$\\
Luminosity $L_{\rm X}$ [erg\,s$^{-1}$] & \multicolumn{3}{c}{$(3.0 \pm 0.2) \times 10^{34}$ } \\
\hline
\hline
\end{tabular}
\label{tab:parnebT2}
\end{center}
\end{table*}

Two temperature plasma models well reproduce the spectra when abundances are allowed to vary. 
These are the first ever spectra of the nebula in \cas, therefore there are no prior constraints 
on nebula abundances. From testing various models, the nitrogen abundance  is not constrained 
and  consistent with the absence of nitrogen. The absolute values of O, Ne, and Mg depend on 
the  initial guess on the C abundance (Table\,\ref{tab:parnebT2}), however, importantly, independently 
on the assumed C-abundance, the relative to oxygen abundances are similar. Strikingly, these ratios 
are different from those derived for the central star (Table\,\ref{tab:parstarT2}).     

\section{Abundances}
\label{sec:aneba}

The abundances determined from the fitting of spectral models to the observed X-ray spectra
of the central star and the nebula can be used to compute the metal mass ratios and fractions. 
We assume that all abundant elements are detected in optical and X-ray spectra. These elements are
C, O, Ne, Mg, Si, and S. Denoting the mass fraction of the $i$-element as $X_{\rm i}$, and normalizing the 
total mass to unity gives $\sum_{\rm i} X_{\rm i} =1$ or 
\begin{equation}
 X_{\rm j}\sum_{\rm i} \frac{X_{\rm i}}{X_{\rm j}} =1
 \label{eq:mass}
\end{equation}
\noindent
The element mass ratios are related to the element
number fractions as 
\begin{equation}
 \frac{X_{\rm i}}{X_{\rm j}}=\frac{n_{\rm i}}{n_{\rm j}}\cdot \frac{A_{\rm i}}{A_{\rm j}}, 
\end{equation}
\noindent
where $n_{\rm i}$ is the abundance by number and $A_{\rm i}$ is the atomic mass of the element $i$.

The parameters of the X-ray spectral models {\em vapec} and {\it bvtapec} which we use to fit the 
spectra with the {\it Xspec} software  are the metal element 
abundances by number relative to their solar values \cite{Asplund2009}. Therefore, 
\begin{equation}
 \frac{X_{\rm i}}{X_{\rm j}}=\frac{n_{\rm i}^{\it Xspec}}{n_{\rm j}^{\it Xspec}}\cdot 
 \frac{n_{\rm i}^{\odot}}{n_{\rm j}^\odot}\cdot
 \frac{A_{\rm i}}{A_{\rm j}},
 \end{equation}
\noindent
then, using Eq.\,({\ref{eq:mass}) and values given in Tables\,\ref{tab:parstarT2} and \ref{tab:parnebT2}, 
one can derive the mass fractions $X_{\rm i}$ as given in Table\,\ref{tab:stpar}. 

From the analysis of the optical spectrum,  $n_{\rm C}=0.25$ and $n_{\rm O}=0.74$, or $n_{\rm C}/n_{\rm O}=0.3$.
Using $n_{\rm i}^{\it Xspec}=n_{\rm i}/n_{\rm i}^\odot$, and calculating input parameters for carbon and oxygen, 
$n_{\rm C}^{\it Xspec}=n_{\rm C}/n_{\rm C}^\odot=929.368$, and $n_{\rm O}^{\it Xspec}=n_{\rm O}/n_{\rm O}^\odot=1510.82$, 
hence as an input ratio in {\it Xspec} $n^{\it Xspec}_{\rm C}/n^{\it Xspec}_{\rm O}=0.615$. 
Solving Eq.\,(\ref{eq:mass}) gives $X_{\rm O}=0.6634$, and then $X_{\rm i}=X_{\rm O}\times\frac{X_{\rm i}}{X_{\rm O}}$.

To estimate the mass of hot gas in the nebula, we make a crude assumption that the nebula consists only of C, O, Ne, 
and Mg. Since X-ray emitting gas is hot, we assume that these ions are fully ionized.  Assuming a constant density $\rho$, 
the spherical nebula mass is $M=\rho V = n_{\rm ion} \mu m_{\rm H} V$, where $V\approx 10^{56}$\,cm$^3$. 
The emission measure of the nebula is $EM=V\mu\beta n_{\rm ion}^2$, where $\beta\approx 1/2$ is the number of electrons 
per atomic mass unit, and $\mu=\sum A_{\rm i}n_{\rm i}/\sum{n_{\rm i}}\approx 13$ is the mean ion mass. 
The $EM$ is observationally constrained, inserting values from Table\,\ref{tab:parnebT2} 
$M=m_{\rm H}\sqrt{V\cdot EM \cdot \mu/\beta} \sim 0.1M_\odot$.
 

\end{document}